# The theory of pinch-effect in the concept scalar-vector potential

F.F. Mende
E – mail: mende@mende.ilt.kharkov.ua

B.I. Verkin Institute for Low Temperature Physics and Engineering,

NAS Ukraine, 47 Lenin Ave., Kharkov, 61164, Ukraiua

## Abstract

In the paper, on the basis of a new concept of the scalar-vector potential introduced by the author, mechanisms resulting in the plasma pinching are considered.

## Introduction

Pitching of plasma when an electric current flows through it, i.e., pinch-effect is important when studying plasma properties and possible plasma applications. Use of pinch-effect is a perspective direction for realization of operated thermonuclear synthesis. Under the description and practical use of this phenomenon there are many works, and we shall not result them, however, the satisfactory theory under the description of this effect is not present.

In all existing works pinch-effect explain within the limits of the concept of a magnetic field. It is supposed, that one current strings create a magnetic field with which other current strings cooperate. Thus, Lorentz's force is directed in such a manner that between these strings takes place the attraction. But in this model there is one absolute obstacle, and it cannot be eliminated neither within the limits of classical electro-dynamics nor within the limits of the special theory of a relativity. It consists that in system of coordinates, which moves together with charges of strings, there are no magnetic fields, therefore in this system of coordinates there are no also forces of an attraction. Other obstacle is that in such scheme of interaction in general participations do not accept ions, and is not clear, what force them to be compressed. Such state of affairs should brake development of works on research and practical use of pinch-effect, in particular within the limits of programs on operated thermonuclear synthesis. In the given paper the alternative theory answering on put questions is offered.

# Model of scalar-vector potential

From the equations of electro-magnetic induction one can obtain transformations of fields for moving coordinate system [1]. If we put $\vec{E}_{||}$ and $\vec{H}_{||}$ for the field components parallel to the velocity direction and $\vec{E}_{\perp}$ and $\vec{H}_{\perp}$ for the perpendicular components, the final fields at the velocity $V$ can be written as

$$\vec{E}'_{||} = \vec{E}_{||},$$

$$\vec{E}'_{\perp} = \vec{E}_{\perp} ch\frac{V}{c} + \frac{Z_0}{V}[\vec{V} \times \vec{H}_{\perp}] sh\frac{V}{c},$$

$$\vec{H}'_{||} = \vec{H}_{||}, \qquad (1)$$

$$\vec{H}'_{\perp} = \vec{H}_{\perp} ch\frac{V}{c} - \frac{1}{Z_0 V}[\vec{V} \times \vec{E}_{\perp}] sh\frac{V}{c},$$

where $Z_0 = \sqrt{\frac{\mu}{\varepsilon}}$ is the space impedance, $c = \sqrt{\frac{1}{\mu\varepsilon}}$ is the velocity of light in the medium under consideration.

As follows from the transformations in Eq. (1) if two charges move at the relative velocity $\vec{V}$, their interaction is determined not only by the absolute values of the charges but by the relative motion velocity as well. The new value of the interaction force is found as

$$\vec{F} = \frac{g_1 g_2 \, ch\frac{V_{\perp}}{c}}{4\pi\varepsilon} \cdot \frac{\vec{r}_{12}}{r_{12}^3}, \qquad (2)$$

where $\vec{r}_{12}$ is the vector connecting the charges, $V_{\perp}$ is the component of the velocity $\vec{V}$, normal to the vector $\vec{r}_{12}$.

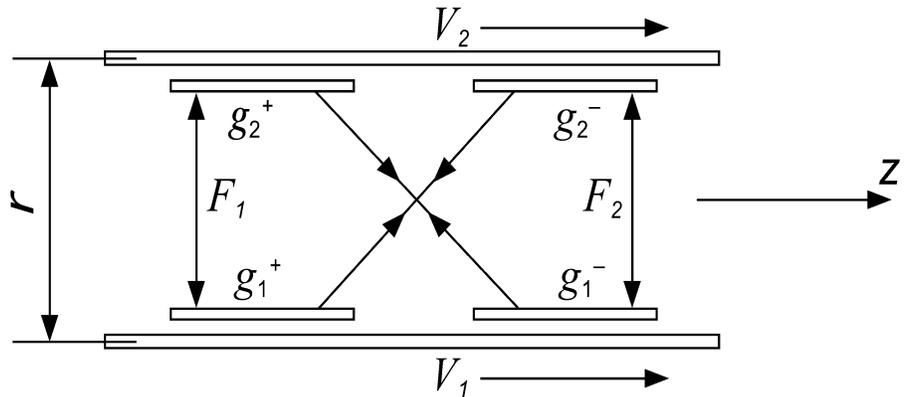

Fig. 1. Schematic view of force interaction between current-carrying wires of a two-conductor line. The lattice is charged positively.

If opposite-sign charges are engaged in the relative motion, their attraction increases. If the charges have the same signs, their repulsion enhances. For $\vec{V} = 0$, Eq. (2) becomes the Coulomb law.

A mew value of the potential $\varphi(r)$ can be introduced at the point, where the charge $g_2$ is located, assuming that $g_2$ is immobile and only $g_1$ executes the relative motion.

$$\varphi(r) = \frac{g_1 \, ch \dfrac{V_\perp}{c}}{4\pi \, \varepsilon \, r} \quad . \tag{3}$$

We can denote this potential as "scalar-vector", because its value is dependent not only on the charge involved but on the value and the direction of its velocity as well.

The interaction of two thin conductors with charges moving at the velocities $V_1$ and $V_2$ (Fig. 1), where $g_1^+$, $g_2^+$ and $g_1^-$, $g_2^-$ are the immobile and moving charges, respectively, pre unit length of the conductors. $g_1^+$ and $g_2^+$ refer to the positively charged lattice in the lower and upper conductors, respectively. Before the charges start moving, both the conductors are assumed to be neutral electrically, i.e. they contain the same number of positive and negative charges.

Each conductor has two systems of unlike charges with the specific densities $g_1^+$, $g_1^-$ and $g_2^+$, $g_2^-$. The charges neutralize each other electrically. To make the analysis of the interaction forces more convenient, in Fig. 1 the systems are separated along the z-axis. The negative-sign subsystems (electrons) have velocities $V_1$ and $V_2$. The force of the interaction between the lower and upper conductors can be considered as a sum of four forces specified in Fig. 1 (the direction is shown by arrows). The attraction forces $F_3$ and $F_4$ are positive, and the repulsion forces $F_1$ and $F_2$ are negative.

According to Eq. (2), the forces between the individual charge subsystems (Fig. 1) are

$$F_1 = -\frac{g_1^+ g_2^+}{2\pi \, \varepsilon \, r} \quad ,$$

$$F_2 = -\frac{g_1^- g_2^-}{2\pi \, \varepsilon \, r} ch\frac{V_1 - V_2}{c} \quad ,$$

$$F_3 = +\frac{g_1^- g_2^+}{2\pi \, \varepsilon \, r} ch\frac{V_1}{c} \quad , \tag{4}$$

$$F_4 = +\frac{g_1^+ g_2^-}{2\pi \, \varepsilon \, r} ch\frac{V_2}{c} \quad .$$

By adding up the four forces and remembering that the product of unlike charges and the product of like charges correspond to the attraction and repulsion forces, respectively, we obtain the total specific force per unit length of the conductor.

Assuming $V \ll c$, we use only the two first terms in the expression of $ch\dfrac{V}{c}$, i.e. $ch\dfrac{V}{c} \cong 1 + \dfrac{1}{2}\dfrac{V^2}{c^2}$, gives

$$F_{\Sigma 1} = \frac{g_1 V_1\, g_2 V_2}{2\pi\, \varepsilon\, c^2 r} = \frac{I_1 I_2}{2\pi\, \varepsilon\, c^2 r}, \tag{5}$$

where $g_1$ and $g_2$ are the absolute values of specific charges, and $V_1$, $V_2$ are taken with their signs.

We see that the forces of interaction between two current-carrying wires are the same as in the concept of magnetic field, however, formation of these forces occurs not only due to moving electrons but also ions of the lattice. In a plasma they are plasma ions. From relations (4) it is seen that for electrons moving at the same velocity with no positively charged lattice, the pinch-effect is absent which is observed in practice.

In practice this method however runs into a serious obstacle. Assuming $g_2^+ = 0$ and $V_2 = 0$, i.e. the interaction, for example, between the lower current-carrying system and the immobile charge $g_2^-$ the interaction force is

$$F_{\Sigma 2} = -\frac{1}{2} \cdot \frac{g_1\, g_2 V_1^2}{2\pi\, \varepsilon\, c^2 r}. \tag{6}$$

This means that the current in the conductor is not electrically neutral, and the electric field

$$E_\perp = \frac{g_1 V_1^2}{4\pi\, \varepsilon\, c^2 r}, \tag{7}$$

is excited around the conductor, which is equivalent to an extra specific static charge on the conductor

$$g = -g_1 \frac{V_1^2}{c^2}. \tag{8}$$

In [1] it is shown that the concentration of electrons in an electron beam moving through an electrically neutral medium is less than that in an immovable beam and is defined by the relation.

$$n_1 = n_0\left(1 - \frac{1}{2}\cdot\frac{V_1^2}{c^2}\right). \tag{9}$$

In this case, the contradiction mentioned above is absent.

Certainly, the given approach is nonconventional, but it is very evident, since it is clearly clear, what forces operate between the charged particles which are carrying out mutual movement. Clearly, that the considered forces will lead to compression a cord of the current, and forces of compression will be more, than forces of pushing apart.

We should note one more circumstance. That density of the electrons which move through a lattice, less, than density motionless of the electrons, specified still F.London [2], however, for electrodynamics and this circumstance of greater consequences had no thermodynamics of superconductors [3], since such amendments are very small.

However, for plasma, especially in case of pinch-effect when density of currents can reach greater sizes, this question gets special value. At greater density of current the difference of longitudinal density of the electrons and ions can reach greater sizes, and it will lead to presence of greater longitudinal electric fields. These fields will try to break off a plasma cord in longitudinal a direction, leading a additional instability about which we earlier did not know. Thus, the new approach opens a way not only to the best physical understanding of pinch-effect, but also predicts the new phenomena which, certainly, it is necessary to consider at research and practical use of pinch-effect.

# References

[1] Mende F.F., On refinement of certain laws of classical electrodynamics, physics/0402084 (http://xxx.lanl.gov/e-print).

[2] London F., Superfluids.Vol.1.Microscopic theory of superconductivity.- Nev York: Dower publ., 1950.- 161p.

[3] F.F. Mende, A.I. Spitsyn, Surface impedance in superconductors, Kiev, Naukova Dumka (1985).- 240p.